# Breast Cancer Molecular Subtypes Prediction on Pathological Images with Discriminative Patch Selecting and Multi-Instance Learning

Hong Liu[1*], Wen-Dong Xu[1,3], Zi-Hao Shang[1,3], Xiang-Dong Wang[1], Hai-Yan Zhou[2], Ke-Wen Ma[2], Huan Zhou[2], Jia-Lin Qi[2], Jia-Rui Jiang[2], Li-Lan Tan[2], Hui-Min Zeng[2], Hui-Juan Cai[2], Kuan-Song Wang[2,4*] and Yue-Liang Qian[1]

[1] Beijing Key Laboratory of Mobile Computing and Pervasive Device, Institute of Computing Technology, Chinese Academy of Sciences, Beijing, China

[2] Department of Pathology, Xiangya Hospital, Central South University, Changsha, Hunan, China

[3] University of Chinese Academy of Sciences, Beijing, China

[4] School of Basic Medical Science, Central South University, Changsha, Hunan, China

**Correspondence:** hliu@ict.ac.cn; wangks001@csu.edu.cn

## Abstract

Molecular subtypes of breast cancer are important references to personalized clinical treatment. For cost and labor savings, only one of the patient's paraffin blocks is usually selected for subsequent immunohistochemistry (IHC) to obtain molecular subtypes. Inevitable block sampling error is risky due to the tumor heterogeneity and could result in a delay in treatment. Molecular subtype prediction from conventional H&E pathological whole slide images (WSI) using AI method is useful and critical to assist pathologists to pre-screen proper paraffin block for IHC. It's a challenging task since only WSI level labels of molecular subtypes from IHC can be obtained without detailed local regions information. Gigapixel WSIs are divided into a huge amount of patches to be computationally feasible for deep learning. While with coarse slide-level labels, patch-based methods may suffer from abundant noise patches, such as folds, overstained regions, or non-tumor tissues. A weakly supervised learning framework based on discriminative patch selecting and multi-instance learning was proposed for breast cancer molecular subtype prediction from H&E WSIs. Firstly, co-teaching strategy using two networks was adopted to learn molecular subtype representations and filter out some noise patches. Then, a balanced sampling strategy was used to handle the imbalance in subtypes in the dataset. In addition, a noise patch filtering algorithm that used local outlier factor based on cluster centers was proposed to further select discriminative patches. Finally, a loss function integrating local patch with global slide constraint information was used to finetune MIL framework on obtained discriminative patches and further improve the prediction performance of molecular subtyping. The experimental results confirmed the effectiveness of the proposed AI method and our models outperformed even senior pathologists, which has the potential to assist pathologists to pre-screen paraffin blocks for IHC in clinic.

# 1 Introduction

Breast cancer is intrinsically heterogeneous and is commonly categorized into molecular subtypes since the late 1990s [1]. According to various molecular expressions of certain genes, breast cancer can be classified into four molecular subtypes including Luminal A, Luminal B, Her-2, and Basal-like [2]. Molecular subtypes directly reveal the biological behavior of breast cancer and represent changes in gene expression, which can be used to determine tailored treatment approaches and predict prognosis [3].

In clinic, molecular subtype diagnosis usually comes from immunohistochemistry (IHC) [4]. IHC uses the high specificity between antigen and antibody, as well as histochemical procedures to mark antigen and antibody positions. IHC staining is used to identify aberrant cells such as those found in cancerous tumors. Certain biological activities, such as growth or cell death, are associated with certain molecular markers [5]. Four biomarkers, including estrogen receptor (ER), progesterone receptor (PR), human epidermal growth factor receptor 2 (HER2), and Ki67, are commonly utilized to immunostain the slides to determine molecular subtypes of breast cancer. Diagnosed subtypes basically determine corresponding treatment strategies, such as targeted drugs for HER2-positive and hormone therapy for Luminal-A. Due to tumor heterogeneity, gene expression of ER, PR, and HER2 often varies in different paraffin blocks and thus may lead to inaccurate subtype diagnosis. For cost and labor savings, pathologists usually examine only one of the paraffin blocks in a case to determine the molecular subtype of breast cancer. Since molecular subtypes determine treatment strategies, inevitable sampling error is risky due to the tumor heterogeneity and could result in a delay in medical treatment. Molecular subtype prediction from conventional H&E pathological whole slide images (WSI) using AI method is useful and critical to assist pathologists to pre-screen proper paraffin block for subsequent IHC in clinic.

Changes in gene expression will cause variations in texture in pathological images. Some pathologists have attempted to investigate the statistical relationship between specific gene expression with Hematoxylin and Eosin (H&E) stained pathological images [6]. Directly predicting molecular subtypes of breast cancer using H&E pathological images based on AI is a prospective study, which may also help improve diagnosis reliability of molecular subtypes.

Molecular subtyping on H&E stained pathological images is a challenging task since we can only obtain the slide-level label for each molecular subtype without detailed local region information. Even experienced pathologists have difficulty annotating corresponding molecular subtype regions in H&E pathological images [7]. Due to the extremely high resolution of whole slide images (WSIs), WSIs are computationally infeasible to be directly fed into a network for training and testing, therefore they are usually divided into small patches. The lack of patch-level labels makes it a weak label problem for machine learning.

Deep learning is becoming increasingly widely used in computer vision tasks. Most deep learning tasks require a large amount of fine-labeled data for supervised learning, which is time-consuming, especially in medical fields. Weakly supervised learning, for example, has been a hotspot for research on reducing the dependence on labeling data. Benenson et al. [8] adopted an interactive method, in which human annotations and the model collaborate to complete the segmentation task. Berthelot et al. [9] augmented labeled data with unlabeled data for classification. To reduce the influence of noisy data, Cheng et al. [10] presented a weakly supervised learning method using a side information network, which largely alleviates the negative impact of noisy image labels. Qu et al.[11] addressed noisy label problem by enforcing prominent feature extraction by matching feature distribution between clean and noisy data.

In recent years, multi-instance learning (MIL) [12] methods are generally adopted for weakly supervised learning. For WSI classification based on MIL, all patches extracted from a pathological image form a bag, and patches are instances of this bag. With only the bag-level labels in the training stage, the goal of MIL is to train a classifier to predict bag-level labels and even instance-level labels. Some previous work extended and enhanced MIL framework using multiple techniques. Wu et al. [13] proposed DE-MIMG that allows each bag to contain pairs of instances and graphs and results in optimal representation. Discriminative bag mapping [14] was adopted to build a discriminative instance pool that can properly separate bags in the mapping space. As attention mechanism gained its popularity in deep neural networks, Ilse et al. [15] and Shi et al. [16] introduced attention mechanism to MIL, where attention weights can represent how much instances contribute to the bag label. Instead of assuming instances in each bag are independent and identically distributed (i.i.d.), Zhang et al. [17] proposed MIVAE that explicitly models the dependencies among instances within each bag for both instance-level and bag-level prediction. Li et al. [18] proposed to use contrast learning to extract multiscale WSI features and a novel MIL aggregator that models the relations of the instances. Shao et al. [19] devised Transformer based correlated MIL that explored both morphological and spatial information. However, most attention-based and correlated MIL methods require large-scale training datasets and significant computational resources. In addition, feature clustering methods have also drawn some attention in MIL. Wang et al. [20] modeled each WSI as k groups of tiles with similar features to ensure learning both diverse and discriminative features. Similarly, Sharma et al. [21] performed K-means clustering on patches within each WSI and randomly sampled a certain amount of patches from each cluster to accommodate for computational limit without much information loss. However, besides the variability of patches within a WSI, the variability of WSIs from the same category is also considerable, where clustering techniques can be used to refine class-level learned features for more accurate subtyping.

Nevertheless, breast cancer molecular subtyping specifically on H&E images has been insufficiently studied. Shamai et al. [22] used logistic regression to explore correlations between histomorphology and biomarker expression and a deep neural network to predict biomarker expression in examined tissue. Rawat et al. [23] introduced "tissue fingerprints" that can learn H&E features to distinguish patients, which are further used to predict ER, PR, and HER2 status. In these studies, machine learning technique is adopted to predict biomarker expression level from H&E histomorphology, direct molecular subtype prediction, however, has not been achieved. Jaber et al. [24] proposed an intrinsic molecular subtype (IMS) classifier from H&E images and analyzed heterogeneity within patches from the same WSI. Although using Inception-v3 to extract features, they adopted traditional PCA and SVM for classification, leading to limited performance.

Since the patches cut from each WSI may come from various regions including lesion, benign, or background of the WSI, some research [25, 26] regard the non-lesion areas in the patches of the pathological images as noisy labels. Differing from pathological classification tasks, such as ductal carcinoma in situ and invasive ductal carcinoma for breast cancer, where pathologists can label tumor regions with different pathological classes, it is impossible to distinguish tumor regions representing different molecular subtypes even for senior pathologists. Although tumor regions annotations are useful information for deep networks to learn molecular subtypes, these manual annotations are time-consuming for pathologists. This paper focuses on molecular subtyping with only slide-level labeling instead of detailed tumor region labeling information. The crucial challenge is to eliminate the influence of noise patches and learn expressive features for classifying molecular subtypes.

In this paper, we modeled the patch-based molecular subtype prediction task of pathological slides as a noisy labeling problem in weakly supervised learning. A multi-instance learning framework DPMIL for pathological image molecular subtyping prediction based on discriminative patch filtering was proposed. First, in order to distinguish noise patches, a pre-classification strategy for molecular classification of pathological slides based on co-teaching was presented. This method adopted co-teaching strategy to train two backbone networks and used co-teaching loss function to filter out noise patches to update model parameters. Then a local outlier factor algorithm was used to reveal the outliers in the feature space for each molecular subtype, and the patches with features close to the cluster center were retained as discriminative patches. Finally, based on the filtered discriminative patches, the pathological slide-level global loss and patch-level local loss were integrated to finetune the prediction model for better feature representation of molecular subtypes. The experimental results confirmed the effectiveness of our proposed framework on the molecular subtyping dataset, breast cancer pathological images provided by Xiangya Hospital. Our AI models outperformed even senior pathologists, which has the potential to assist in pre-screening proper paraffin block of patients for subsequent IHC molecular subtyping in clinic.

## 2 Materials and Methods

### 2.1 Data, Software and Hardware

This paper used breast cancer H&E pathology dataset BCMT (Breast Cancer with Molecular Typing) provided by Xiangya Hospital. All the pathology WSIs used a pyramid storage structure.

As Table 1 shows, the BCMT dataset contains 1254 pathological WSIs from 1254 patients or cases with slide-level molecular subtype annotations between 2017 and 2019. The dataset contains 313 slides for Luminal A, 382 slides for Luminal B, 316 slides for Her-2 overexpression subtype, and 243 slides for the Basal-like subtype. We randomly divided the slides into training set and validation set with a ratio of 8:2 for each type. This paper uses accuracy, precision, recall, and F1 score to measure the performance of four molecular subtypes.

We use 4 GeForce GTX 2080 Tis with 11GB memory to train the network and Python with Pytorch to implement our algorithm. The initial learning rate is 0.1 and the poly learning rate policy with the power of 0.9 is employed. The minibatch size is set as 32.

### 2.2 Proposed Framework

This paper proposes a breast cancer molecular subtype prediction framework based on multi-Instance learning and discriminative patch filtering. The pipeline of our framework is illustrated in Figure 1.

**Table 1.** Distribution of each molecular subtyping in BCMT dataset.

| Set   | Luminal A | Luminal B | Her-2 | Basal-like | Total |
|-------|-----------|-----------|-------|------------|-------|
| Train | 254       | 298       | 255   | 196        | 1003  |
| Val   | 59        | 84        | 61    | 47         | 251   |
| Total | 313       | 382       | 316   | 243        | 1254  |

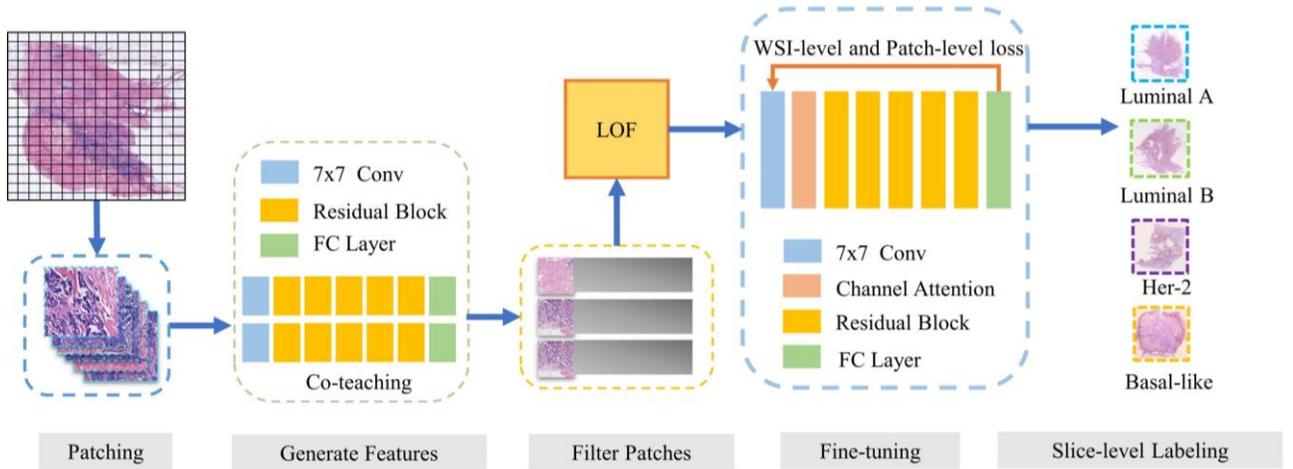

The pipeline of our framework contains 5 stages. (1) WSIs are divided into patches. (2) ResNet trained with Co-teaching generate feature for each patch. (3) LOF is adopted to select discriminative patches based on features. (4) Discriminative patches are used to finetune ResNet with WSI and patch loss. (5) Finetuned model predict final molecular subtypes for patches and WSIs.

**Figure 1.** Our framework DPMIL for molecular subtype prediction.

Firstly, patches from H&E WSIs are extracted to train a molecular subtype classifier. Co-teaching [27] between two networks is used to obtain the patch-level classification and select candidate discriminative patches. Then local outlier factor (LOF) [28] based on cluster centers of subtypes is adopted to further filter out noise patches and obtain discriminative patches. Based on these discriminative patches, we finetuned a new molecular subtyping model initialized by the model performed better in co-teaching stage. Finally, the local loss function and global loss function are combined as constraint information in multi-instance learning framework to improve feature representation of molecular subtypes. The finetuned model is used to obtain the final patch-level and slide-level molecular subtyping results.

### 2.3 Feature Construction and Patch Selection based on Co-teaching

In multi-instance learning framework, each patch is usually assigned the same label as WSI it belongs to [29, 30, 31]. While for molecular subtyping, patches from WSI may contain benign or other tissues, which will make slide-level prediction difficult. To reduce these noise patches, this paper adopts co-teaching strategy [27], which usually trains two neural networks and enables them to learn from each other. This strategy assumes that the two models simultaneously consider the samples with the lowest loss as non-noisy samples. These selected instances are considered more representative of the category of the bag than other instances. Each network treats samples with minimal loss in each batch as knowledge and feeds these samples to the other network. Co-teaching strategy is inherently suitable for classification with noisy labels.

This paper uses ResNet-50 [32] as the backbone for co-teaching. The parameters of the two models are randomly initialized and the selection strategy of $K$ follows [27]. During co-teaching process, the ResNet-50 network is used to obtain representative features and confidence for each patch. Patches with higher confidence are selected as candidate discrimination patches for subsequent process.

### 2.4 Noise Patch Filtering Using Local Outlier Factor

Although the above co-teaching strategy used co-teaching loss to filter out some noise patches, many noise patches from benign or other tissue regions remain. For selected high confidence patches, we can

MIL based breast cancer subtyping

obtain the feature of each patch before the classification layer. Patches belonging to the same molecular subtype tend to gather into the same cluster in feature space.

This paper further proposed a noise filtering method based on local outlier factors (LOF), which is a classic density-based algorithm [28]. The main idea is to calculate a numerical score to represent the abnormality degree of a sample to the cluster center with average density. In feature space, the density of a certain point is compared with the average density of points around it. If the former score is lower, the point may be abnormal and vice versa.

Figure 2 shows an example of point set (blue point) in feature space for certain molecular subtyping. We query whether these four points are outliers of the point set. The green point is not an outlier with a lower LOF score, and the red points are outliers with high ones. The size of the red point is the value of the LOF scores and represents the abnormality degree of a certain point.

We perform LOF for each subtype of molecular features and regard patches that do not belong to a specific cluster of molecular subtype as noise patches.

### 2.5 Multi-Instance Learning with Global and Local Constraint

The above selected discriminative patches are further used to improve feature representation of molecular subtypes based on multi-instance learning framework (MIL). MIL regards the WSI as a bag containing a number of patches. These patches are considered as instances and their predictions are aggregated to obtain a bag-level prediction. ResNet-50 is also adopted as backbone to train MIL classification model. We initialize the MIL model with the model that performs better in co-teaching and use discriminative patches for finetuning.

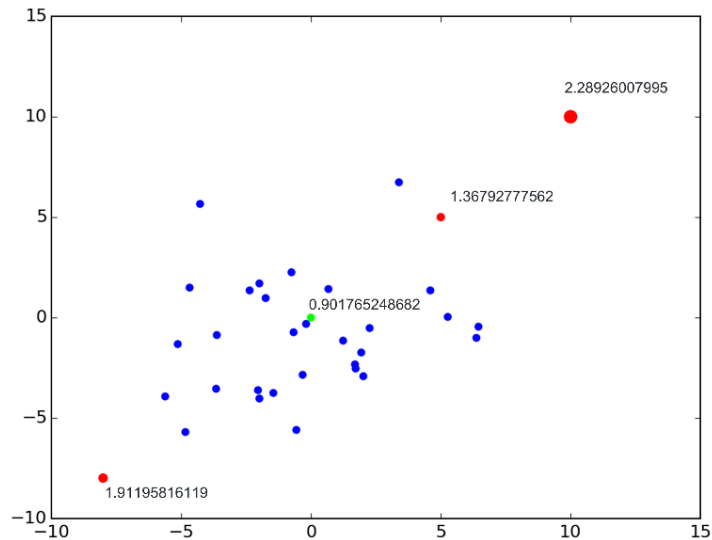

**Figure 2.** Local outlier factor example.

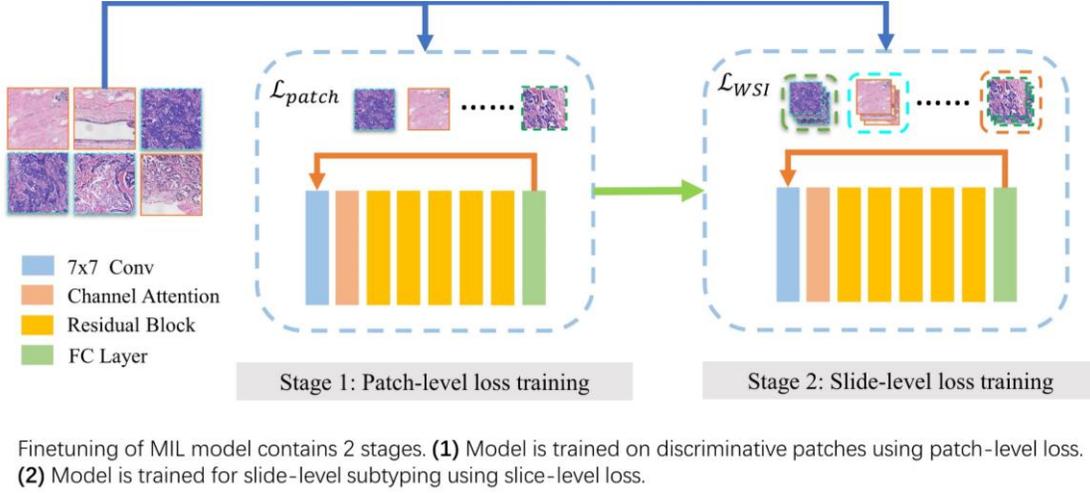

Finetuning of MIL model contains 2 stages. **(1)** Model is trained on discriminative patches using patch-level loss. **(2)** Model is trained for slide-level subtyping using slice-level loss.

**Figure 3.** Two-stage training of MIL model

We introduce the slide-level loss function to impose global information constraints to guide the MIL training. The slide-level loss function $L_{WSI}$ is defined as Formula 1, where $L_{WSI_i}$ represents the slide-level loss function of the i-th pathological image defined as the cross-entropy function [32]. $N_{WSI}$ represents the total number of pathological slides in the training set, and α is the weight of slide-level loss.

$$L_{WSI} = \alpha \frac{1}{N_{WSI}} \sum_{i=1}^{N_{WSI}} L_{WSI_i} \qquad (1)$$

$L_{WSI_i}$ is defined as Formula 2, where M is the molecular type number, and $y_{o,c}$ is the indicator function. When the output prediction result in o is the same as the true label c of the pathological slide, it is set to 1, otherwise, it is 0.

$$L_{WSI} = -\sum_{c=1}^{M} y_{o,c} \, log(P_c) \qquad (2)$$

$P_c$ is defined in Formula 3, representing the confidence level of slide-level molecular subtyping. $N_p$ is the total number of patches of the pathological image, $p_{i,c}$ represents the confidence value when the i-th patch of WSI is classified as type $c$. As shown in Formula 3, the average confidence value of all patches from the same WSI are obtained and used as the slide-level molecular subtyping confidence.

$$P_C = \frac{1}{N_p} \sum_{i=1}^{N_p} p_{i,c} \qquad (3)$$

The two-stage training diagram is shown in Figure 3. We use the patches by LOF-Denoising as input. In each epoch, the training process is divided into two stages. The first stage uses all patches to calculate the patch-level loss to train the model, and we use cross-entropy as the loss function, which is defined in Formula 4.

$$L_{CE} = -\sum_{c=1}^{M} y_c \log(p_c) \qquad (4)$$

$M$ represents the total number of molecular types and $y_c$ is the indicator function, which equals 1 when prediction $c$ equals the ground truth of the slide. $p_c$ denotes the confidence level the current patch is classified as type $c$. The second stage is trained for slide-level subtyping using slide-level loss function as global constraint information.

## 3    Results

This section introduces several experiments to evaluate the performance of our proposed framework DPMIL, including the patch resampling strategy, co-teaching, LOF, and MIL training successively. The performance of model is evaluated using average accuracy, recall, precision, and macro F1 for four subtypes.

### 3.1    Results of Patch Resampling and Co-teaching

The total number of different types of patches at different resolutions is shown in Figure 4, which shows the imbalance of number of patches for each molecular subtype and each resolution. To deal with the imbalance of dataset, we use a patch resampling strategy to ensure category equalization. For each epoch, the number of training data for each molecular subtype is set as a constant value. The common part is randomly sampled from all patches, and the number of sampled patches is different according to their resolution: 180,000 patches at 5X, 700,000 patches at 10X and, 5,000,000 patches at 20X. The rare part of the data is generated by data augmentation such as randomly flip, horizontal, and vertical symmetry.

We use ResNet-50 as the classifier to evaluate the performance of the sampling strategy. Figure 5 shows the results of molecular subtyping with patches resampling at different resolutions. The accuracy of models with resampling strategy are all higher than those without resampling at three resolutions.

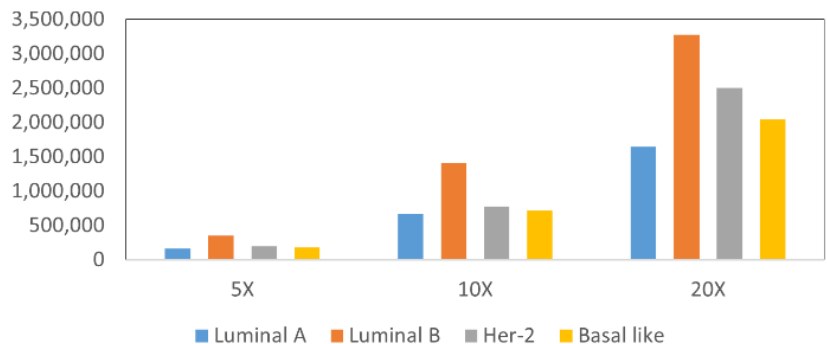

**Figure 4.** Statistics patches of different molecular subtypes at each resolution.

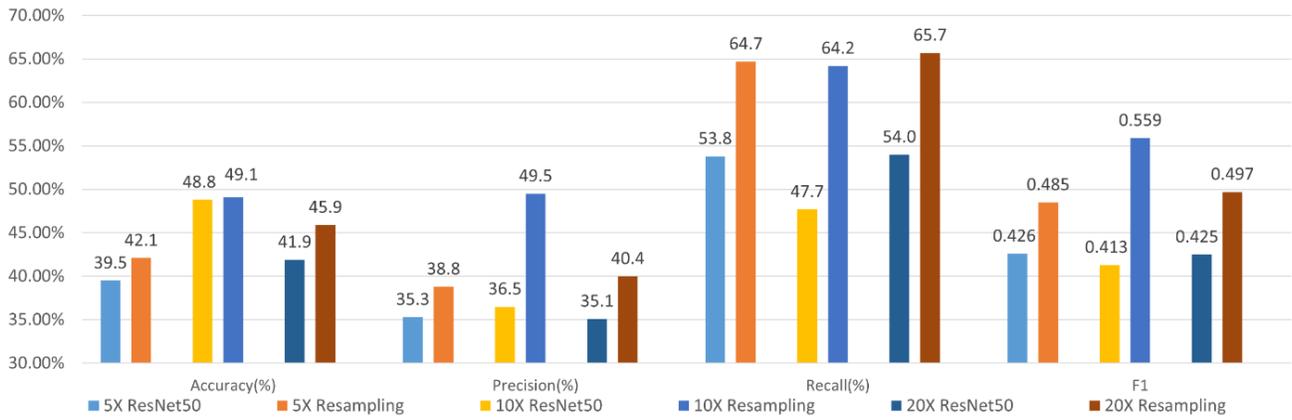

**Figure 5.** Results of 4-class molecular subtype classification with patch resampling at different resolutions.

And F1 values improve about 6% with patch resampling methods for all the resolutions. In addition, the highest accuracy and F1 value are all achieved at 10X, which indicates that patch size and tissue texture make a good compromise at 10X.

### 3.2 Molecular Subtype Classification Using Co-teaching and LOF

This section describes experiments to verify the effectiveness of the co-teaching strategy. ResNet-50 was selected as two backbones for co-teaching. The model is trained for 20 epochs with a minibatch of 32. The initial value of the learning rate is 0.01, and the polynomial learning rate decay method [34] is used to adjust the learning rate.

Figure 6 shows the results of molecular subtype classification with and without co-teaching at different resolutions. The accuracy improves 4% to 6% and F1 score improves 4% to 11% with co-teaching. The co-teaching framework trains two neural networks and enable them to learn from each other, which can reduce the influence of noise patches. The F1 value of 10X-Co-teaching reaches 0.604 and improves 4.5% compared with 10X-resampling.

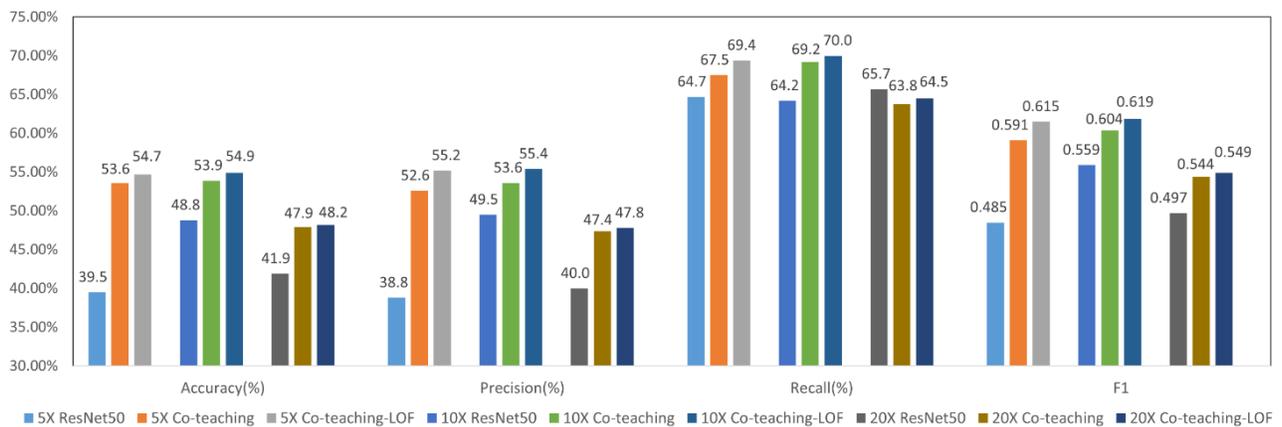

**Figure 6.** Results of 4-class molecular subtype classification with co-teaching and LOF at different resolutions.

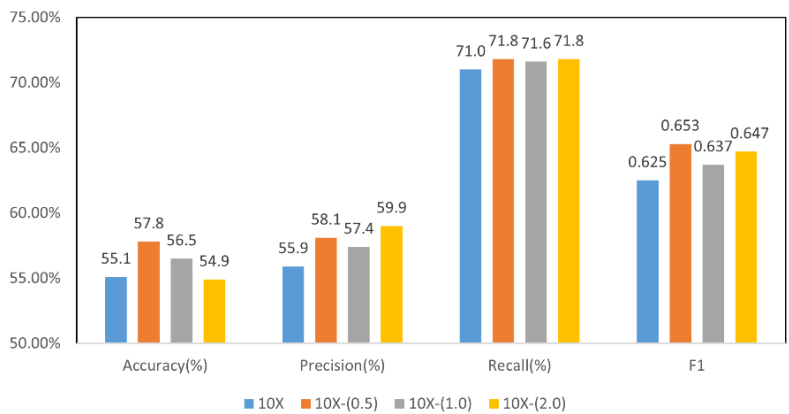

**Figure 7.** Results of 4-class molecular subtype classification with MIL finetuning at different resolutions.

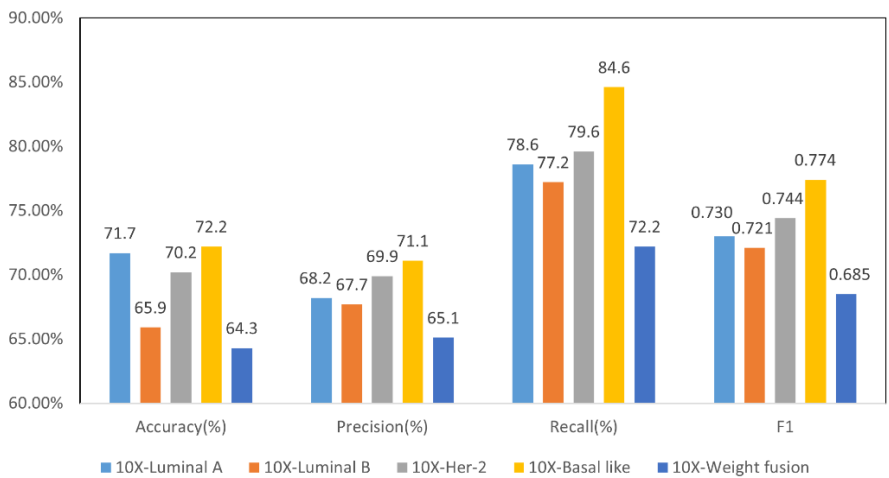

**Figure 8.** Results of 2-class molecular subtype classification and 4-class weighted fusion at 10X resolution.

We selected the model from Co-teaching with the higher F1 value at each resolution. Features before classification layer were input into LOF-Denoising for patch filtering for all molecular types. We supposed $S_i$ is the number of normal patches of i-th molecular type and there were $\sum_{i=1}^{4} S_i$ features in total. These features in co-teaching were used for statistical classification of output logits, the number of which is limited to 2000. The experimental results are shown in Figure 6, which shows LOF after co-teaching can further improve the metrics since more noise patches are filtered out. We select 10X resolution in the following experiments.

### 3.3 Multi-Instance Learning with Global Information

Based on the above discriminative patch selection, we further verify the multi-instance learning framework with slide-level loss. We used a four-class classification model for molecular subtyping and compared the results with different weights in Formula 1. In the second training stage of the model with global constraint, the influence of the weight $\alpha$ in loss function of formula 1 was examined.

When $0 \leq \alpha \leq 1$, the influence of the second stage on the model parameters is weakened. When $\alpha = 0$, the second stage of training does not affect the model. When $\alpha > 1$, the influence of the second stage is enhanced. We set the value to 0.5, 1.0, and 2.0 respectively to evaluate the effectiveness of global loss constraint in the second stage of training.

Figure 7 shows the results of MIL for molecular subtyping, proving that using slide-level loss function can improve the performance of the model. The reason may be that there are still some noise patches in the selected patches after noise filtering. We used a slide-level loss to add global constraint information, which can further reduce the influence of noise patches.

### 3.4 Binary classification model and weighted fusion

Apart from the four-class classification model, to further improve the performance of molecular subtype classification, we also tried binary classification models for each molecular subtype. Finally, a weighted fusion method is adopted to accomplish the final four-type classification.

Binary classification models were trained similar to four-class classification model, including co-teaching, LOF, and slide-level loss of MIL. Parameter $\alpha$ is set to 0.5 for all the experiments. The prediction results of each molecular type of binary classification model are shown in Figure 8, where F1 reached over 0.72 for all subtypes. Notably, Basal-like molecular type obtained the highest F1 value of 0.774.

For four-class classification, we averaged the confidence level of all patches from a WSI, and then use it as the confidence of the molecular subtype of the WSI. We used grid search [35] for the best weight setting of the four-class prediction model and finally take 0.6, 0.9, 0.5 and 0.7 as weights of four subtypes. The final weighted four-class classification results are shown in Figure 8. Compared with the direct four-class molecular type prediction from model 10X- (0.5) in Figure 7, four classifiers weighted fusion in Figure 8 can increase the accuracy by 6.7% and the F1 score by 3.2%.

To compare our method with pathology doctors in molecular type classification, nine pathologists were invited to diagnose molecular subtypes of a total of 99 randomly selected WSIs from test dataset. In clinic, pathologists usually can classify molecular subtypes on IHC images but not on H&E stained images. Therefore, pathologists can only conduct subtyping totally based on image pattern and their clinical experience. Table 2 shows the average accuracy, precision, recall and macro F1 scores according to the labels pathologists (D1: 5 years' experience, D2: 10 years' experience, D3: 15 years' experience) assigned to each H&E WSI. Specifically, we provide the means and ranges of 4 metrics from seven 5-year pathologists (D1s). As shown in Table 2, 5-year experienced doctors can hardly make better predictions than random guess, which indicates the unusual difficulty in breast cancer subtyping on H&E images. To be optimistic, more experienced doctors can provide a more accurate diagnosis on molecular subtypes. Our four-class classification model (10X-0.5) and fused binary classification model (10X-Weight fusion) show obvious superiority over doctors in all metrics, surpassing predictions of the most experienced doctor (D3) by 15.4% and 21.5% in accuracy and F1 respectively.

Table 2. Comparison of molecular subtyping results among doctors and our best models.

| Predictor | Accuracy | Precision | Recall | F1 |
|---|---|---|---|---|
| Mean of D1s | 28.8% | 28.4% | 28.9% | 0.260 |
| Range of D1s | 23.2~32.3% | 23.2~33.6% | 25.2~32.4% | 0.202~0.286 |
| D2 | 38.4% | 40.2% | 39.0% | 0.394 |
| D3 | 42.4% | 42.5% | 43.9% | 0.429 |
| 10X- (0.5) (Ours) | 57.8% | 58.1% | 71.8% | 0.653 |
| 10X-Weight fusion (Ours) | 64.3% | 65.1% | 72.2% | 0.685 |

## 4    Discussion

Molecular subtyping is becoming more and more important in the therapy of malignant disease. But accurate molecular subtyping on H&E images is challenging due to tumor heterogeneity. Pathologists should examine every paraffin block of the tumor in order to confirm subtype theoretically. But it is so costly that most pathologists usually examine only one block in a case. In this situation, the sampling error is inevitable and a predictive AI model for molecular subtyping on H&E images can significantly improve the present clinical procedure. Pathologists can quickly make preliminary subtype predictions of a tumor and select the most representative block based on our AI model. Then the representative block is examined to further confirm the molecular subtyping prediction with IHC in clinic. In addition, the inference time and computing resources our model requires are negligible compared to the expensive IHC. Therefore, pathologists could avoid the sampling error with the help of AI model at almost no additional cost and provide a more reliable result for oncologists to improve curative effect.

Since WSI-level labels lack detailed region annotation information, most of the existing methods use patch-based methods for WSI recognition. How to eliminate the influence of noise patches and learn the corresponding features for molecular subtyping through training process is the key problem. Our work aims to predict slide-level labels of H&E pathological slides using only weakly annotated information at slide-level. This paper proposes a framework by selecting these discriminant patches to reduce the impact of noise patches and combined MIL for molecular subtype classification. The experimental results show the effectiveness of our proposed framework on the partner hospital's breast cancer H&E pathological image dataset.

MIL has been applied in diverse diseases and image modalities including classification of cancer in histopathology images, dementia in brain MR, tuberculosis in X-ray images and others. MIL classifiers can benefit from information about cooccurrence and structure of instances when classifying bags [36]. For example, Melendez et al. [37] trained a MIL classifier only with X-ray images labeled as healthy or abnormal, yet outperforming its supervised version trained on outlines of tuberculosis lesions.

Some studies combine traditional machine learning algorithms with weakly supervised learning and apply them to pathological slide classification tasks. Hou et al. [38] combined the EM method based on multi-instance learning with a convolutional neural network and used it to predict patch-level results. Campanella et al. [39] used a recurrent neural network model to extract feature representations between different patch examples to obtain a slide-level classification for basal cell carcinoma and breast cancer axillary lymph node metastasis. Raju et al. [40] proposed a graph attention clustering multi-instance learning algorithm based on texture features to predict the TNM staging of rectal cancer tumor metastasis and improved the accuracy of pathological slide staging. Wang et al. [41] proposed a classification framework for pathological slides for gastric cancer diagnosis, which used localization networks to extract patch features and critical filtered patches to replace the general clustering module. After local network extraction and screening of key patch feature maps, concatenation is performed to obtain an overall feature map describing pathological slides.

Recent studies rely largely on the powerful feature extraction capability of deep learning. Yang et al. [42] trained a six-type classifier for identification of lung lesions from WSIs based on EfficientNet [43]. To obtain slide-level diagnosis, a thresh-old-based tumor-first aggregation method that fused majority voting and probability threshold was proposed. Wang et al. [44] developed a second-order multiple instances learning method with an adaptive aggregator stacked by attention mechanism and RNN for histopathological image classification, attempting to explore second-order statistics of deep features for histopathological images. MIL framework can also be applied to similar tasks like survival prediction. Yao et al. [45] proposed Deep Attention Multiple Instance Learning by introducing Siamese MI-FCN that learns features from phenotype clusters, and attention-based MIL pooling that performs trainable weighted aggregation. While our paper focuses on the selection of discriminative patches and combined local and global constraint information in MIL framework.

The retrospective study design would have resulted in inevitable bias and all the data were collected from a single center, thereby limiting the sample size of the study. In future work, we will combine multi-center and multi-resolution information of pathological images to improve the accuracy and to evaluate on larger datasets.

## 5    Conclusions

Molecular subtype prediction from H&E pathological slides is a challenging task. Based on slide-level weak labels, this paper proposes a multi-instance learning framework for molecular subtype classification with discriminative patches selection. Firstly, we use co-teaching strategy to train the molecular subtype prediction model with noise patches. Then, the noise patches are filtered out according to features obtained from the model through local outlier factor algorithm. Finally, based on the filtered discriminative patches, a multi-instance learning based molecular subtyping model using both slide-level and patch-level loss is finetuned. The experimental results show the effectiveness of proposed framework on breast cancer H&E pathological image dataset from Xiangya hospital. Although its performance is not sufficient to replace pathologists' clinical diagnosis directly, it is reasonable to employ our framework to preliminary screening for more convenient and reliable molecular subtyping.

**Funding**

This work was supported by the Beijing Natural Science Foundation (Z190020) and National Natural Science Foundation of China (81972490).

**References**

1.   Todd R Golub, Donna K Slonim, Pablo Tamayo, Christine Huard, Michelle Gaasenbeek, Jill P Mesirov, Hilary Coller, Mignon L Loh, James R Downing, Mark A Caligiuri, et al. Molecular classification of cancer: class discovery and class prediction by gene expression monitoring. science, 286(5439):531–537, 1999
2.   Lajos Pusztai, Chafika Mazouni, Keith Anderson, Yun Wu, and W Fraser Symmans. Molecular classification of breast cancer: limitations and potential. The oncologist, 11(8):868–877, 2006.
3.   Yersal O, Barutca S. Biological subtypes of breast cancer: Prognostic and therapeutic implications[J]. World journal of clinical oncology, 2014, 5(3): 412.
4.   Sengal A T, Haj-Mukhtar N S, Elhaj A M, et al. Immunohistochemistry defined subtypes of breast cancer in 678 Sudanese and Eritrean women; hospitals based case series[J]. BMC cancer, 2017, 17(1): 1-9.
5.   Whiteside G, Munglani R. TUNEL, Hoechst and immunohistochemistry triple-labelling: an improved method for detection of apoptosis in tissue sections—an update[J]. Brain Research Protocols, 1998, 3(1): 52-53.
6.   Zhu X, Yao J, Zhu F, et al. Wsisa: Making survival prediction from whole slide histopathological images[C]. Proceedings of the IEEE Conference on Computer Vision and Pattern Recognition. 2017: 7234-7242.
7.   Chen J M, Li Y, Xu J, et al. Computer-aided prognosis on breast cancer with hematoxylin and eosin histopathology images: A review[J]. Tumor Biology, 2017, 39(3): 1010428317694550.
8.   Benenson R, Popov S, Ferrari V. Large-scale interactive object segmentation with human annotators[C] //Proceedings of the IEEE/CVF Conference on Computer Vision and Pattern Recognition. 2019: 11700-11709.
9.   Berthelot D, Carlini N, Goodfellow I, et al. Mixmatch: A holistic approach to semi-supervised learning[J]. Advances in Neural Information Processing Systems, 2019, 32.

10. Cheng L, Zhou X, Zhao L, et al. Weakly supervised learning with side information for noisy labeled images[C]//European Conference on Computer Vision. Springer, Cham, 2020: 306-321.
11. Qu Y, Mo S, Niu J. DAT: Training Deep Networks Robust To Label-Noise by Matching the Feature Distributions[C]//Proceedings of the IEEE/CVF Conference on Computer Vision and Pattern Recognition. 2021: 6821-6829.
12. Yao J, Zhu X, Jonnagaddala J, et al. Whole slide images based cancer survival prediction using attention guided deep multiple instance learning networks[J]. Medical Image Analysis, 2020, 65: 101789.
13. Wu J, Zhu X, Zhang C, et al. Multi-instance multi-graph dual embedding learning[C]//2013 IEEE 13th International Conference on Data Mining. IEEE, 2013: 827-836.
14. Wu J, Pan S, Zhu X, et al. Multi-instance learning with discriminative bag mapping[J]. IEEE Transactions on Knowledge and Data Engineering, 2018, 30(6): 1065-1080.
15. Ilse M, Tomczak J, Welling M. Attention-based deep multiple instance learning[C]//International conference on machine learning. PMLR, 2018: 2127-2136.
16. Shi X, Xing F, Xie Y, et al. Loss-based attention for deep multiple instance learning[C]//Proceedings of the AAAI Conference on Artificial Intelligence. 2020, 34(04): 5742-5749.
17. Zhang W. Non-IID Multi-Instance Learning for Predicting Instance and Bag Labels using Variational Auto-Encoder[J]. arXiv preprint arXiv:2105.01276, 2021.
18. Li B, Li Y, Eliceiri K W. Dual-stream multiple instance learning network for whole slide image classification with self-supervised contrastive learning[C]//Proceedings of the IEEE/CVF Conference on Computer Vision and Pattern Recognition. 2021: 14318-14328.
19. Shao Z, Bian H, Chen Y, et al. Transmil: Transformer based correlated multiple instance learning for whole slide image classification[J]. Advances in Neural Information Processing Systems, 2021, 34.
20. Wang X, Chen H, Gan C, et al. Weakly supervised deep learning for whole slide lung cancer image analysis[J]. IEEE transactions on cybernetics, 2019, 50(9): 3950-3962.
21. Sharma Y, Shrivastava A, Ehsan L, et al. Cluster-to-conquer: A framework for end-to-end multi-instance learning for whole slide image classification[C]//Medical Imaging with Deep Learning. PMLR, 2021: 682-698.
22. Shamai G, Binenbaum Y, Slossberg R, et al. Artificial intelligence algorithms to assess hormonal status from tissue microarrays in patients with breast cancer[J]. JAMA network open, 2019, 2(7): e197700-e197700.
23. Rawat R R, Ortega I, Roy P, et al. Deep learned tissue "fingerprints" classify breast cancers by ER/PR/Her2 status from H&E images[J]. Scientific reports, 2020, 10(1): 1-13.
24. Jaber M I, Song B, Taylor C, et al. A deep learning image-based intrinsic molecular subtype classifier of breast tumors reveals tumor heterogeneity that may affect survival[J]. Breast Cancer Research, 2020, 22(1): 1-10.
25. Karimi D, Dou H, Warfield S K, et al. Deep learning with noisy labels: Exploring techniques and remedies in medical image analysis[J]. Medical Image Analysis, 2020, 65: 101759.
26. Xue C, Dou Q, Shi X, et al. Robust learning at noisy labeled medical images: Applied to skin lesion classification[C]//2019 IEEE 16th International Symposium on Biomedical Imaging (ISBI 2019). IEEE, 2019: 1280-1283.
27. Han B, Yao Q, Yu X, et al. Co-teaching: Robust training of deep neural networks with extremely noisy labels[J]. Advances in neural information processing systems, 2018, 31.
28. Breunig M M, Kriegel H P, Ng R T, et al. LOF: identifying density-based local outliers[C]//Proceedings of the 2000 ACM SIGMOD international conference on Management of data. 2000: 93-104.
29. Chikontwe P, Kim M, Nam S J, et al. Multiple instance learning with center embeddings for histopathology classification[C]//International Conference on Medical Image Computing and Computer-Assisted Intervention. Springer, Cham, 2020: 519-528.
30. Hashimoto N, Fukushima D, Koga R, et al. Multi-scale domain-adversarial multiple-instance CNN for cancer subtype classification with unannotated histopathological images[C]//Proceedings of the IEEE/CVF conference on computer vision and pattern recognition. 2020: 3852-3861.
31. Chetan L Srinidhi, Ozan Ciga, and Anne L Martel. Deep neural network models for computational histopathology: A survey. Medical Image Analysis, page 101813, 2020.
32. He K, Zhang X, Ren S, et al. Deep residual learning for image recognition[C]//Proceedings of the IEEE conference on computer vision and pattern recognition. 2016: 770-778.
33. Goodfellow I, Bengio Y, Courville A. Deep learning[M]. MIT press, 2016.

34. He T, Zhang Z, Zhang H, et al. Bag of tricks for image classification with convolutional neural networks[C]//Proceedings of the IEEE/CVF Conference on Computer Vision and Pattern Recognition. 2019: 558-567.
35. Chicco D. Ten quick tips for machine learning in computational biology[J]. BioData mining, 2017, 10(1): 1-17.
36. Carbonneau M A, Cheplygina V, Granger E, et al. Multiple instance learning: A survey of problem characteristics and applications[J]. Pattern Recognition, 2018, 77: 329-353.
37. Melendez J, van Ginneken B, Maduskar P, et al. A novel multiple-instance learning-based approach to computer-aided detection of tuberculosis on chest x-rays[J]. IEEE transactions on medical imaging, 2014, 34(1): 179-192.
38. Hou L, Samaras D, Kurc T M, et al. Patch-based convolutional neural network for whole slide tissue image classification[C]//Proceedings of the IEEE conference on computer vision and pattern recognition. 2016: 2424-2433.
39. Campanella G, Hanna M G, Geneslaw L, et al. Clinical-grade computational pathology using weakly supervised deep learning on whole slide images[J]. Nature medicine, 2019, 25(8): 1301-1309.
40. Raju A, Yao J, Haq M M H, et al. Graph attention multi-instance learning for accurate colorectal cancer staging[C]//International Conference on Medical Image Computing and Computer-Assisted Intervention. Springer, Cham, 2020: 529-539.
41. Wang S, Zhu Y, Yu L, et al. RMDL: Recalibrated multi-instance deep learning for whole slide gastric image classification[J]. Medical image analysis, 2019, 58: 101549.
42. Yang H, Chen L, Cheng Z, et al. Deep learning-based six-type classifier for lung cancer and mimics from histopathological whole slide images: a retrospective study[J]. BMC medicine, 2021, 19(1): 1-14.
43. Tan M, Le Q. Efficientnet: Rethinking model scaling for convolutional neural networks[C]//International Conference on Machine Learning. PMLR, 2019: 6105-6114.
44. Wang Q, Zou Y, Zhang J, et al. Second-order multi-instance learning model for whole slide image classification[J]. Physics in Medicine & Biology, 2021.
45. Yao J, Zhu X, Jonnagaddala J, et al. Whole slide images based cancer survival prediction using attention guided deep multiple instance learning networks[J]. Medical Image Analysis, 2020, 65: 101789.